\documentclass[twoside]{ilcws07}
\usepackage[latin1]{inputenc}
\usepackage[dvips]{graphicx,epsfig,color}
\usepackage{wrapfig,rotating}
\usepackage{amssymb,amsmath,array}

\begin{document}
\title{Simulation Studies and Detector Scenarios \\for an ILC Polarimeter}
\author{K.Oleg Eyser$^1$, Christian Helebrant$^{1,2}$, Daniela K\"afer$^1$, Jenny List$^1$ and Ulrich Velte$^{1,3}$
\vspace{.3cm}\\
1 - DESY Hamburg, Germany
\vspace{.1cm}\\
2 - Universit\"at Hamburg, Germany
\vspace{.1cm}\\
3 - Leibniz Universit\"at Hannover, Germany
}

\maketitle

\begin{abstract}
  High energy longitudinal electron polarimetry will be based on Compton
  scattering for the International Linear Collider.
  An unobtrusive measurement has to include a magnet chicane setup serving
  as a spectrometer.
  Current proposals make use of Cerenkov detectors for electron detection.
  A fast simulation has been developed to study the basic properties of
  scenarios for the polarimeter setup.
\end{abstract}

\section{Introduction}
Electron polarimetry at the International Linear Collider (ILC) 
will be based on Compton scattering using a laser colliding head-on
with the incident lepton beam in the beam delivery system.
The shape of the differential Compton cross section depends strongly
on the helicity states of electron beam and laser.
Asymmetries can then be determined from different helicity 
configurations, which scale with the longitudinal polarization of the
electron bunches \cite{TESLA}.

\begin{figure}[bh]
  \centerline{\includegraphics[width=\columnwidth]{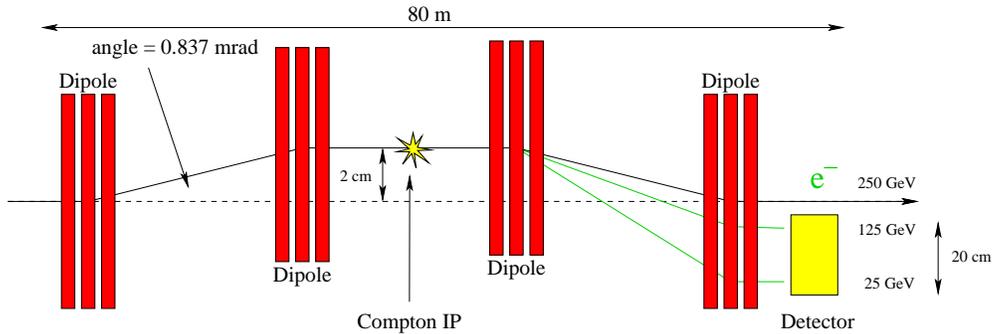}}
  \caption{\label{fig_chicane}
    Proposed layout of the upstream electron polarimeter in a magnet chicane.
  }
\end{figure}

At ILC energies of 250 GeV or more, the angular distributions of both 
the scattered photons and the recoil electrons are constrained to within 
a few $\mu$rad for all but the most highly energetic photons.
The polarimeter, therefore, will be placed in a magnet chicane serving
as a spectrometer for the recoil electrons (see figure \ref{fig_chicane}).
The setup consists of four sets of dipoles in such a way that the emittance
of the incident electron beam should not be impaired by more than 1\%.
The electrons are first displaced parallel to their original direction by 
about 2~cm to the side by two sets of dipoles.
After that, a pulsed laser (10~ps with 35~$\mu$J) hits the bunches head-on
and deflects about $10^3$ out of the $20\cdot10^9$ electrons.
The second pair of dipole sets puts the undisturbed electrons back on their
original track and serves as a spectrometer for the recoil particles which
are deflected more strongly due to their reduced energy.

The Compton electron yield will then be determined after the last dipole set
in a Cerenkov detector.
The detector needs to cover less than 20~cm horizontally and only a
fraction of that in vertical direction due to the orientation of the dipoles.
The overall length of the polarimeter is reaching almost 80~m to accomplish
this spread and to comply with the accelerator emittance demands at the same
time.
Cerenkov light will be produced by the electrons in gas tubes or quartz fibers,
alternatively.
The light is then transferred further to photo-detectors which are placed out 
of the accelerator plane in order to reduce background from beam related
interaction or other effects from the beam delivery system.

The polarimeter location and setup have to be chosen carefully, since the
polarization needs to be known as close to the electron-positron interaction 
point as possible.
Especially the direction of the polarization vector can change easily in the
magnetic fields of the beam delivery system.
Beam orbit alignment accuracies have been estimated to be smaller than 
80~$\mu$rad at 250~GeV for the measurement to be in compliance with the 
physics demands for the polarization determination 
($\Delta P/ P \leq 0.25\%$, \cite{TESLA,Positron}).
The polarimeter setup is such that the Compton edge does not change its
position in the detector with respect to the incident beam energy.
This is especially of value during commissioning of the accelerator in order 
to keep track of polarization preservation and also in threshold scans.

\section{Fast Simulations}
A fast Monte Carlo simulation has been developed which is used for first
analyses of the basic properties of the polarimeter layout and the detector
design.
Bunches of $20\cdot10^9$ electrons at 250~GeV and 80\% polarization are 
coupled to the circularly polarized, green laser pulses (2.33~eV) by a
Compton generator.
Both beams have an assumed circularly distributed width with a gaussian shape
of $\sigma$ = 50~$\mu$m.
Also, there is a slight crossing angle of 10~mrad which is necessary in the
experimental setup later.

The Compton recoil electrons are tracked through the dipoles and translated
into Cerenkov photons at the location of the detector through a refraction
index of 1.0014 for $C_{4}F_{10}$ or pressurized propane (1.1~atm).
Light transmission to the photo-detectors is currently covered by a global
efficieny of 55\%.

The light yield detected by the photo-detectors is determined from the 
wavelength dependent quantum efficiency.
It has been taken from conventional {\sc Hamamatsu R6094} photo-multiplier
tubes.
Between 300~nm $\leq\lambda\leq$ 650~nm, the quantum efficiency reaches a
maximum of 25\% around 350~nm and drops to zero elsewhere due to the opacity 
of the entrance window glass.
{\sc Adc}-counts are finally derived from the detected photons.
These counts can be distorted by a quadratic function that describes the
general behaviour of differential non-linearities in the electronics 
read-out chain.
The distortion is maximal for medium values and vanishes at zero counts and 
when reaching saturation level.

\begin{figure}[tbh]
\centerline{\includegraphics[width=\columnwidth]{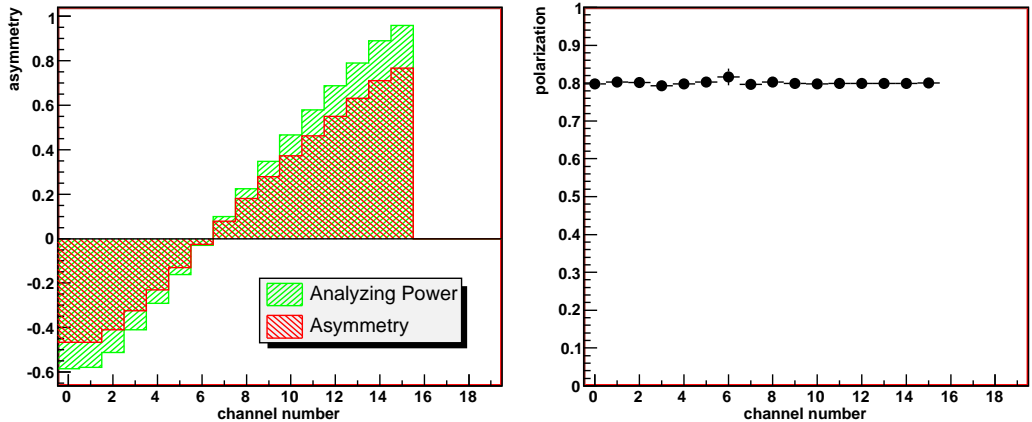}}
  \caption{\label{fig_asymmetry}
    Asymmetry $\epsilon$ and analyzing power $A$ as functions of channel number.
    The ratio of both leads to the electron longitudinal polarization $P$.
    }
\end{figure}

The {\sc Adc}-counts are determined for same and opposite helicity
configurations and accumulated over several bunches.
From these, an asymmetry is calculated as a function of channel number, 
see left side of figure \ref{fig_asymmetry}.
The channel number reflects the transverse position of the Compton electrons
at the detector surface (2~cm $\leq x\leq$ 20~cm) and is, therefore, 
a function of the inverse recoil energy.

The polarization $P$ is calculated from the {\sl measured} asymmetry $\epsilon$ 
and the calculated analyzing power $A$, i.e., the Compton asymmetry for 
completely polarized electron bunches:
\begin{equation}
  \epsilon = A\cdot P = \frac{N^{+}_{\sc Adc}-N^{-}_{\sc Adc}}{N^{+}_{\sc Adc}+N^{-}_{\sc Adc}},
\end{equation}
where $N^{+}_{\sc Adc}$ and $N^{-}_{\sc Adc}$ are the integrated {\sc Adc}-counts.

Polarizations are determined for each channel seperately and then combined
to a weighted mean (see right side of figure \ref{fig_asymmetry}).
This way, problematic channels can be excluded from the average value,
which is especially important for the zero crossing of the asymmetry 
(channel 6 in figure \ref{fig_asymmetry}).
Statistical errors diminish after accumulation over many bunch trains
of each 2820 bunches and reveal systematic uncertainties.
\begin{figure}[bht]
\centerline{\includegraphics[width=\columnwidth]{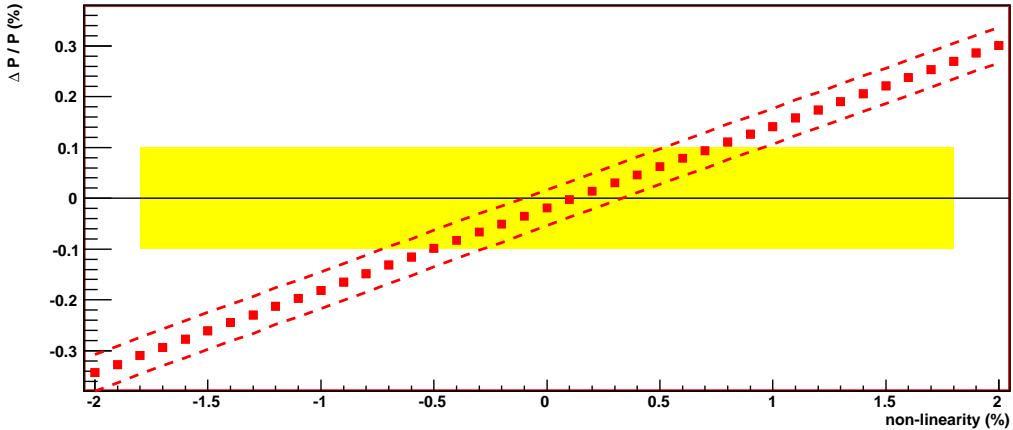}}
  \caption{\label{fig_nonlin}
    Effect of electronics differential non-linearities on the 
    determination of the electron polarization.
    The yellow box represents the range of the desired maximum contribution
    to the total error $\Delta P/P\leq0.25\%$.
    }
\end{figure}

Experimentally, the polarization will be extracted from the output
{\sc Adc}-counts without proper knowledge of integral or differential
non-linearities.
In the simulation, one can compare the experimental values with the primary
Compton electrons or with the secondary Cerenkov photons, so the effect of
the distorted {\sc Adc}-readout becomes traceable for single channels.
Also, the measured polarization can be checked with respect to the {\sl real}
(input) electron polarization.
Figure \ref{fig_nonlin} shows the deviations between measured and real
polarizations as a function of the quadratic non-linearities for a 20-channel
detector.

In order to meet the physics demand for the polarization measurement, it 
is necessary to be in control of differential non-linearities in the range 
of 0.5\%, contributing less than 0.1\% to $\Delta P/P$.
This effect is decreasing with increasing channel numbers.
However, for Cerenkov tubes, a channel width of 1~cm seems to be a reasonable
size.
First non-linearity measurements have been carried out in a test stand
\cite{LCWS:Kaefer} and more detailed studies of the whole detector setup 
are planned.

\section{Outlook}
The simulation so far is very limited and only includes non-linearities of
the electronics read-out after tracking the Compton electrons towards the
detector surface and transforming them into Cerenkov photons.
In the future, a full picture of the polarimeter has to include other details
of the magnet chicane and the detector setup.
This includes a proper description of the gas (or possibly other) volumes, the
geometry for the transmission with reflectivities and absorption, and light
extraction in the photo-detectors.
There are on-going efforts to use {\sc Bdsim}\footnote{Beam Delivery Simulation}
\cite{BDSIM} for a description of the magnet chicane and add an accurate 
{\sc Geant4} model of the complete detector setup.
This simulation would also serve as a comprehensive tool to study the
possibilities of combining the beam emittance measurement or others with the
polarimeter chicane.
Such an additional measurement might induce additional downstream background
which has to be considered carefully.

\section{Acknowledgments}
The authors acknowledge the support by DFG Li 1560/1-1.

\begin{footnotesize}

\end{footnotesize}
\end{document}